\documentstyle[preprint,prl,aps]{revtex}
\input psfig
\begin{document}
\draft
%\documentstyle{europhys} 
%\input euromacr.tex
%\newcommand{\text}[1]{\hbox{\rm #1}} 
%\renewcommand{\vec}[1]{\mbox{\boldmath ${#1}$}} 
%\renewcommand{\thefootnote}{\fnsymbol{footnote}}  
%\euro{}{}{}{} 
%\Date{} 
%\shorttitle{O. Entin-Wohlman et al. Transmission of 2 interacting 
%electrons}
\title{Transmission of two interacting electrons}
\author{O. Entin-Wohlman$^1$, Amnon Aharony$^1$, Y. Imry$^2$
 and Y. Levinson$^2$}
%\institute
\address{$^1$~{\it School of Physics and Astronomy, Raymond and Beverly 
Sackler 
Faculty of Exact Sciences, 
Tel Aviv University, Tel Aviv 69978, Israel \\}
$^2$~{\it Department of Condensed Matter Physics, The Weizmann 
Institute of Science,
Rehovot 76100, Israel}}
%\rec{}{in final form }

%\pacs{
%\Pacs{73}{23.-b}{Mesoscopic systems}
%\Pacs{72}{10.-d}{Theory of electronic transport; scattering mechanisms}
%}

\maketitle
\begin{abstract}
The transmission of two electrons through a region where they interact
is found to be enhanced by a renormalization of the repulsive
interaction.
For a specific example of the single--particle
Hamiltonian, which includes a strongly attractive potential,
the renormalized interaction becomes attractive,
and the transmission has a pronounced maximum as function of the depth of
the single--electron attractive potential.
The results apply directly to a simple model of scattering of
two interacting
electrons by a quantum dot.
%, with consequences for the Coulomb blockade picture.

\end{abstract}

%\section{Introduction}

Transport through quantum dots, \cite{kastner}
systems of quantum dots, \cite{kouwenhoven}
or one--dimensional wires \cite{ness} is currently the subject of much
interest. Such structures have potential applications as 
artificial atomic or molecular
devices, and are very instrumental in studying
strong correlations and their implications, e.g., the Kondo effect \cite
{goldhaber} or magnetic transitions. \cite{su}
The theoretical studies of the transmission through such systems 
concentrate mostly on the single--electron conductance,
\cite{ng} 
with the electron--electron interaction taken into account 
in an approximate manner, e.g., via the charging energy
(introducing the notion of the dot capacitance \cite{CB}) 
or by using the slave--boson
mean--field approximation. \cite{georges} Exact solutions of
interacting electrons, in this context, focus mainly on the electronic 
spectrum of 
two electrons, confined to a restricted region in space, which models the 
quantum
dot (see, e.g., Ref. \cite{wagner}). Obviously, two is the minimal
number of electrons required to study 
interactions. However, devices including two electrons
can be investigated experimentally, \cite{su,kane} and may have potential
applications in quantum computers. \cite{kane1}
Recently, we have studied the electronic spectrum of two interacting 
electrons on
artifical atoms, \cite{aharony}
and found that the understanding of the full physical behavior
of such systems requires a consideration 
of the {\it entire} system, {\it including
the leads}.
In that calculation we found that {\it interactions can delocalize 
one of the electrons},
depending on the strength of the connections between the dot and the leads.
In other words, it is not
sufficient to treat the interacting region as an isolated system. 
 This transition of an electron from the bound state into the band may
be relevant for the metal-insulator transition brought about by donor
ionization.
The 
exact solution of the two electron problem yields a renormalized value of 
the
interaction, which depends on the energy.

In this paper we study the one--dimensional
transmission of two electrons, when they 
interact via a
contact potential at a certain point.
We consider the Hamiltonian
\begin{eqnarray}
{\cal H}(x_{1},\sigma_{1},x_{2},\sigma_{2})&=&{\cal H}_{0}(x_{1},x_{2})
+U\delta (x_{1})\delta (x_{2})\delta_{\sigma_{1},-\sigma_{2}},\nonumber\\
{\cal H}_{0}(x_{1},x_{2})&=&{\cal H}_{\rm sp}(x_{1})+{\cal H}_{\rm 
sp}(x_{2}),
\label{H}
\end{eqnarray}
where $x_{i}$  and $\sigma_{i}$ are the coordinate and the spin component 
of 
the $i$th electron, and
${\cal H}_{\rm sp}$ is the single--particle Hamiltonian,
independent of the spin components.
In (\ref{H}), $U$ is the local (Hubbard) interaction.
This is a continuum version of the Anderson impurity model.
\cite{anderson}
Since ${\cal H}_{0}$ does not depend on the spins, and since the 
interaction
vanishes on the
triplet states (which are antisymmetric in space and vanish when
$x_1=x_2=0$), it is convenient to separate the Hilbert space into triplet
and singlet states. The former are unaffected by the interaction, and
their scattering is fully described by the single--electron 
non--interacting
transmission. The rest of this paper is devoted to an exact solution
of the singlet case.
We express the scattered wave function in terms of the single--particle
spectrum, and calculate the {\it two--electron} current.

Our explicit calculations are carried out for a single--particle 
Hamiltonian which
has an attractive short-range potential, resulting with a
 bound state, whose inverse localization length $\kappa$
determines the electric current
(or equivalently the transmission) in the absence 
of the
interaction. When one views the model as a `quantum dot' coupled to ideal 
one--dimensional
conductors (`leads'), then the strength of the attractive 
single--particle potential can be
regarded as the gate voltage applied to the `dot'. 
\cite{kastner,kouwenhoven} 

We consider explicitly two
scenarios for the transmitted current. In the first case,
the incident wave contains
two propagating
electrons (with wave numbers $p_1$ and $p_2$).
For our contact interaction [Eq. (1)], we
show that in this case there is no effect of the
interaction: for a sample of 
large length $L$, the current is given by
\begin{equation}
j=e(p_{1}|t_{p_{1}}|^{2}+p_{2}|t_{p_{2}}|^{2})/(mL)+O(U/L^2),
\label{p1p2}
\end{equation}
where $t_p$ is the single--electron non--interacting transmission
amplitude.
The reason for
this simple behavior is quite clear: In our model,
the two electrons ``feel" one another
only when both are at $x=0$, and the amplitude for this to occur
is proportional to $1/L$.

In the second case,
the ``impinging" wave contains one propagating electron (with wave number $p$),
while the other is captured by the
attractive potential.
Now the current is found to have the form 
\begin{equation}
j=e p T(U,p)/(mL),
\label{j2}
\end{equation}
and the effective transmission $T(U,p)$ has a very interesting dependence 
on
the single--electron parameters and on $U$. Specifically, Fig. \ref{fig11}
shows the dependence of $T(U,p)$ on $\kappa$, for several values of
$U$ and $p$ (in units of $\hbar^2/(2m)$ and of the large wave number 
cutoff $\omega$, 
respectively).
Clearly, $T(U,p)$
increases significantly when $U>0$ (compared to the case $U=0$),
reflecting the ``screening" of the attractive single-electron potential by 
the bound electron, eventually leading to a transition from the doubly 
bound `insulating' state to a state in which one electron is `free', which
can be called `metallic'.\cite{aharony}
For the special $\delta-$function attractive
potential discussed below, and for large $\kappa$,
the arguments of Ref. \cite{aharony} imply that
this transition happens for $U>2$.
Indeed, when $U<2$
Fig. \ref{fig11} (left panels) shows a monotonic decrease in $T$, but $T$ is
much larger
than the non--interacting result. At $U=2$ (central panels), 
$T$ is found to approach a 
finite plateau
value even at large $\kappa$. This value increases with $p$.
For $U>2$, we observe a peak in $T$, which moves to larger values of
$\kappa$ as $U$ approaches 2 from above. This peak has $T \approx 1$ for
small $p$, but it decreases and broadens as $p$ increases.
As we show below, this peak corresponds to the resonance with the 
just-bound doubly-occupied  state.
$T(U,p)$ decreases for larger $\kappa$, when the effective renormalized
interaction becomes more and more attractive.

In addition to the above peak (and to the peak at $\kappa=0$),
we sometimes find a third peak at $p \sim \kappa \ll \omega$,
i. e. when the total energy of the two electrons vanishes; at larger
$\kappa$'s the incoming electron is no longer able to `ionize' the bound
electron (see e. g. $T(4,.1)$). 
At this point, both the product of the transmitted and reflected 
wave functions and the product of the `bound' and `free' electron
wave functions on the `dot' are maximal, giving a peak in the interaction
between the free and the bound electron and therefore in the screening.
This peak disappears as $p$ increases.
The peaked structure of the transmission versus $\kappa$, which represents
the peaks of $T$ as function of the gate voltage on the `dot', is very
different from that found in the naive Coulomb blockade picture. \cite{CB}
In the latter case, the distances between peaks would be equal to $U$!
We conclude that this naive picture fails for our exactly
solved example.

We now give more technical details.We consider only
the singlet scattering wave functions,
which are spatially  {\it symmetric},
$\Psi(x_{1},x_{2})=\Psi(x_2,x_1)$.
At total energy $E$, we split $\Psi$ into
%\begin{equation}
$\Psi
=\Psi_{0}+
\Psi_{\rm S}$,
%\label{Psi}
%\end{equation}
where $\Psi_{0}$ is the `incoming' solution of ${\cal H}_{0}$, 
with the same energy $E^+\equiv (E+i\eta )$ (with $\eta \rightarrow 0^+$),
%\begin{equation}
$({\cal H} _{0}-E^+)\Psi_{0}=0$.
%\end{equation}
It then follows that 
\begin{equation}
\Psi_{\rm S}(x_{1},x_{2})=
UG_{E}(x_{1},x_{2};
0,0)
\Psi_{0}(0,0),
\label{Psis}
\end{equation}
where $G_{E}$ is the two--particle Green's function of the {\it interacting
Hamiltonian}, obeying 
%\begin{eqnarray}
\begin{equation}
({\cal H}-E^+)G_{E}(x_{1},x_{2};
x_{1}',x_{2}')=
%\nonumber\\
-\delta (x_{1}-x_{1}')\delta (x_{2}-x_{2}')
.\label{HG}
%\end{eqnarray}
\end{equation}

For the model Hamiltonian given by (\ref{H}) one can express 
%the Green's function
%of the interacting system, 
$G_{E}$ in terms of the two--particle 
Green's function
of the {\it non--interacting} system,
%. Denoting the latter by 
$G_{E}^{0}$:
%, we have
%\begin{eqnarray}
\begin{equation}
({\cal H}_{0}-E^+)G_{E}^{0}(x_{1},x_{2};
x_{1}',x_{2}')=
%\nonumber\\
-\delta (x_{1}-x_{1}')\delta (x_{2}-x_{2}')
.\label{H0G0}
%\end{eqnarray}
\end{equation}
Combining Eqs. (\ref{HG}) and (\ref{H0G0}) yields
%\begin{eqnarray}
\begin{equation}
G_{E}(x_{1},x_{2};
x_{1}',x_{2}')=G_{E}^{0}(x_{1},x_{2};x_{1}',x_{2}')
%\nonumber\\
%&&~~~
+UG_{E}^{0}(0,0;x_{1}',x_{2}')
G_{E}(x_{1},x_{2};
0,0),
%\end{eqnarray}
\end{equation}
and hence
\begin{equation}
G_{E}(x_{1},x_{2};0,
0)=
G_{E}^{0}(x_{1},x_{2};0,0)/[1-UG_{E}^{0}(0,0;0,0)] 
\equiv F_E G_{E}^{0}(x_{1},x_{2};0,0)/U.
\label{GE00}
\end{equation}
This immediately yields $G_{E}(x_{1},x_{2};x_{1}',
x_{2}')$ and
%so that
%%\begin{eqnarray}
%\begin{equation}
%G_{E}(x_{1},x_{2};x_{1}',
%x_{2}')=G_{E}^{0}(x_{1},x_{2};x_{1}',x_{2}')
%%\nonumber\\
%&&~~~~~~
%+\frac{G_{E}^{0}(x_{1},x_{2};0,0)G_{E}^{0}(0,0;x_{1}',x_{2}')}
%{1-UG_{E}^{0}(0,0;0,0)}.\label{GE}
%%\end{eqnarray}
%\end{equation}
%
%The singlet wave function of the two electrons,
%at energy $E$, is thus given by 
%\begin{eqnarray}
\begin{equation}
\Psi
(x_{1},x_{2})=\Psi_{0}(x_{1},x_{2})
%\nonumber\\
%&&~~~~~
+F_E G_{E}^{0}(x_{1},x_{2};0,0)\Psi_{0}(0,0).
\label{fPsi}
\end{equation}
%\end{eqnarray}
%where 
%\begin{equation}
%F_{E}=U/[1-UG_{E}^{0}(0,0;0,0)],\label{FE}
%\end{equation}
%and where we have used Eqs. (\ref{Psis}) and (\ref{GE00}).
The right hand side of ({\ref{fPsi}) is determined
solely by the eigenstates of the
non--interacting Hamiltonian ${\cal H}_{0}$. 
%One notes that the
%interaction 
$U$ appears only in the form $F_E$,
which turns out to be 
quite important in
determining the transmission characteristics.

Instead of solving directly for the two--electron non--interacting
Green's function $G_{E}^{0}$, we present this function in terms of
the single--particle Green's function, $g_{\epsilon}(x,x')$, 
of ${\cal H}_{\rm sp}$,
\begin{equation}
\Bigl ({\cal H}_{\rm sp}(x)-\epsilon^+\Bigr )g_{\epsilon}(x,x')=
-\delta (x-x'),
\end{equation}
where $\epsilon^+=\epsilon+i \eta$, and $\eta \rightarrow 0^{+}$. 
%
%This Green's function 
$g_{\epsilon}(x,x')$ can also be
written in terms of its spectral decomposition,
\begin{equation}
g_{\epsilon}(x,x')=\sum_{n}\frac{\phi_{n}(x)\phi_{n}^{\ast}(x')}{\epsilon^+
-\epsilon_{n}},
\end{equation}
%Here $\epsilon_{n}$ and $\phi_{n}$ are the eigenenergies and eigenstates 
where ${\cal H}_{\rm sp} \phi_{n}= \epsilon_{n} \phi_{n}$. 
Writing the non--interacting
singlet spatial wave functions
as 
\begin{equation}
\Psi_0^{nm}(x_1,x_2)=\Bigl (\phi_{n}(x_{1})\phi_{m}(x_{2})+
\phi_{n}(x_{2})\phi_{m}(x_{1})\Bigr )/2^{(1+\delta_{nm})/2},
\label{psinm}
\end{equation}
with energy $E(n,m)=\epsilon_n+\epsilon_m$, one can show that
the spectral decomposition of the singlet non--interacting 
two--particle
Green's function is
%\begin{eqnarray}
\begin{equation}
G_{E}^{0}(x_{1},x_{2};x_{1}',x_{2}')=
%\nonumber\\
%&&~~~
\sum_{nm}
\frac{\phi_{n}(x_{1})\phi_{m}(x_{2})\phi_{n}^{\ast}(x_{1}')\phi_{m}^{\ast}(
x_{2}')}
{E^+ -\epsilon_{n}-\epsilon_{m}}.\label{GE0}
%\end{eqnarray}
\end{equation}
Using the two spectral representations, we obtain
\begin{eqnarray}
&&G_{E}^{0}(x_{1},x_{2};x_{1}',x_{2}')=
%\nonumber\\
-\frac{1}{\pi}\int_{-\infty}^{\infty}
~d \epsilon ~g_{E-\epsilon}(x_{1},x_{1}')\Im g_{\epsilon}(x_{2},x_{2}')=
\nonumber\\
&&\frac{i}{2\pi}\int_{-\infty}^{\infty}~d\epsilon
~g_{E-\epsilon}(x_{1},x_{1}')g_{\epsilon}
(x_{2},x_{2}'),\label{fGE0}
\end{eqnarray}
where the last equality follows from the Kramers--Kronig relations.

We next calculate 
the quantum average of the current density operator at $x=x_0$,
%\cite{mahan}
%\begin{equation}
%\hat{j}(x_{0})=\frac{e}{2im}\sum_{i}\Bigl (\delta
%(x_{i}-x_{0})\frac{d}{dx_{i}}+\frac{d}{dx_{i}}\delta (x_{i}-x_{0})\Bigr ),
%\end{equation}
%$\hat{j}(x_{0})=\hat{j}_1+\hat{j}_2$,
in the exact singlet state $\Psi$. Noting that 
$\Psi (x_{1},x_{2})$ is symmetric in
$x_{1}$, $x_{2}$, this average is
\begin{equation}
j(x_{0})=\frac{2e\hbar}{m}
\Im \int dx_{1}dx_{2}\delta (x_{1}-x_{0})
\Psi^{\ast}(x_{1},x_{2})\frac{d}{dx_{1}}\Psi (x_{1},x_{2}),
%\frac{e\hbar}{im}\int~dx_{1}dx_{2}\delta (x_{1}-x_{0})\times\nonumber\\
%&&\Bigl (\Psi_{0}^{\ast}(x_{1},x_{2})+F_{E}^{\ast}\Psi_{0}^{\ast}(0,0)
%(G^{0}_{E}(x_{1},x_{2};0,0))^{\ast}\Bigr )\times\nonumber\\
%&&\frac{d}{dx_{1}}
%\Bigl (\Psi_{0}(x_{1},x_{2})+F_{E}\Psi_{0}(0,0)
%G_{E}^{0}(x_{1},x_{2};0,0)\Bigr )+cc.
\end{equation}
where $\Psi$ is given by Eq. (\ref{fPsi}).
The explicit calculation of $j$ now requires only integrals involving
the non--interacting functions $\Psi_0(x_1,x_2)$ and $g_\epsilon(x,0)$.
In what follows, we shall assume that $\Psi_0 \equiv \Psi_0^{pq}$, as 
given by
Eq. (\ref{psinm}), and that the total energy is given by 
$E=\epsilon_p+\epsilon_q$.
The calculation is then facilitated using identities such as
\begin{eqnarray}
&&\int~dx \phi_p^\ast(x)g_{\epsilon}(x,0)=
\phi_p^\ast(0)/(\epsilon^+-\epsilon_{p})
;\nonumber\\
&&\int~dxg_{\epsilon_{1}}^{\ast}(x,0)g_{\epsilon_{2}}(x,0)=
\frac{g_{\epsilon_{1}}^{\ast}
(0,0)-g_{\epsilon_{2}}(0,0)}{\epsilon_{2}-\epsilon_{1}+i\eta};\nonumber\\
&&\int~d\epsilon g_{E-\epsilon}(x,0)/(\epsilon^+-\epsilon_{p})
=-2\pi ig_{E-\epsilon_{p}}(x,0),
\label{formulae}
\end{eqnarray}
where the last equation represents the Kramers--Kronig relations.

To proceed, we need to specify ${\cal H}_{\rm sp}$. As the simplest
possible example, we choose
% for which we will calculate 
%the transmission explicitly, 
a simple $\delta$--function attractive
potential,
\begin{eqnarray}
{\cal H}_{\rm sp}(x)=-\frac{\hbar^{2}}{2m}\frac{d^{2}}{dx^{2}} 
-V\delta (x),\label{Hsp}
\end{eqnarray}
which has one bound state, $\phi_b=\sqrt{\kappa}e^{-\kappa |x|}$
with the inverse localization length $\kappa=mV/\hbar^2$, and
with eigenenergy $-\epsilon_b=-\hbar^2\kappa^2/(2m)$,
and ``band" scattering wave functions
$\phi_p=(e^{i p x}+r_p e^{i p |x|})/\sqrt{L}$, with the
reflection and transmission amplitudes
\begin{equation}
r_p=i \kappa/(p-i \kappa),\ \ \ t_p=p/(p-i \kappa),
\label{ref}
\end{equation}
and with eigenenergy
$\epsilon_p=\hbar^2p^2/(2m)$.
For this simple Hamiltonian one has \cite{economu}
\begin{eqnarray}
g_{\epsilon}(x,x')=g_{\epsilon}(x',x)=
-im\Bigl (
e^{ik_{\epsilon}|x-x'|}+
r_{k_\epsilon}e^{ik_{\epsilon}(|x|+|x'|)}\Bigr )/(\hbar^{2}k_{\epsilon}),
\label{g}
\end{eqnarray}
where the wave vector $k_{\epsilon}$ is defined by
%\begin{equation}
$k_{\epsilon}=\sqrt{2m\epsilon^+}/\hbar$, with 
$\Im k_{\epsilon}>0$.
%\end{equation}

There are two physical
situations which are of interest for
the non--interacting wave function $\Psi_0^{pq}$.
The first 
%occurs when $\Psi_{0}$
corresponds to two {\it propagating} electrons, impinging from the left,
when both $p=p_1$ and $q=p_2$ represent ``band" states.
%with total energy $E(p_{1},p_{2})=\epsilon_{p_{1}}+\epsilon_{p_{2}}$.
%&&\frac{1}{\sqrt{2}L}\Bigl [\Bigl
%(e^{ip_{1}x_{1}}+r_{p_{1}}e^{ip_{1}|x_{1}|}
%\Bigr )\Bigl (e^{ip_{2}x_{2}}+r_{p_{2}}e^{ip_{2}|x_{2}|}
%\Bigr )\nonumber\\
%&&~+x_{1}\leftrightarrow x_{2}\Bigr
%],~~E(p_{1},p_{2})
%=\epsilon_{p_{1}}+\epsilon_{p_{2}},~~\epsilon_{p}=\frac{\hbar^{2}p^{2}}{2m
%},
%\end{eqnarray}
%where $L$ is the normalization volume. Here $r_{p}$ ($t_{p}$)
%is the reflection (transmission) amplitude of
%the single--particle Hamiltonian, 
%\begin{equation}
%r_{p}=\frac{i\kappa}{p-i\kappa},~~~t_{p}=\frac{p}{p-i\kappa}.
%\end{equation}
This yields the simple Eq. (\ref{p1p2}).
We devote the rest of this paper to 
the second, more interesting, choice for $\Psi_0^{pq}$, 
when one electron is propagating, with $p$ representing its wave vector,
and the
other is captured in the bound state ``$b$", i. e. $q=i\kappa$.
Now the total energy is
%\begin{equation}
$E(p,b)=\epsilon_{p}-\epsilon_{b} \equiv \hbar^2(p^2-\kappa^2)/(2m)
\ge -\epsilon_b$.
%,\label{Epb}
%\end{equation}
%and can also be negative, when $\epsilon p<\epsilon_{b}$. 
%Here $\Psi_{0}$ is
%\begin{eqnarray}
%&&\Psi_{0}^{p,b}(x_{1},x_{2})=\nonumber\\
%&&\sqrt{\frac{\kappa}{2L}}\Bigl [e^{-\kappa |x_{1}|}
%\Bigl (e^{ipx_{2}}+r_{p}e^{ip|x_{2}|}\Bigr )
%+x_{1}\leftrightarrow x_{2}\Bigr ]
%\end{eqnarray}
In this case, the terms coming from the interaction are of the same order
as the non--interacting ones, as now 
the amplitude
for the two electrons to be together at $x=0$ is of order $\sqrt{\kappa 
/L}$. 
A long calculation, using Eqs. (\ref{GE0}),
(\ref{fGE0}), (\ref{formulae}) and (\ref{g}), now yields
Eq. (\ref{j2}),
and the effective transmission is
given by
\begin{equation}
T(U,p)=|t_{p}|^2-2m\Re(F_E r_p t_p)/\hbar^2.
\label{T}
\end{equation}
%From Eq. (\ref{FE}) it is clear that t
The second term here will have
a ``resonance" (yielding the large--$\kappa$ peak in Fig. 1)
when the real part of the denominator of $F_E$ (cf. 
Eq. (\ref{GE00}))
will vanish, i. e. when   
\begin{equation}
U_{\rm eff}(E)^{-1}\equiv \Re (F_E^{-1})
=U^{-1}-\Re G_{E}^{0}(0,0;0,0)\label{Uef}
\end{equation}
vanishes. In some sense, $U_{\rm eff}(E)$ represents the renormalized
interaction.
The details of the transmission thus require explicit expresions
for
$G_{E}^{0}(0,0;0,0)$. 
Introducing an upper cutoff 
$W=\hbar^{2}\omega^{2}/2m$
on the ``band" states, and using 
Eq. (\ref{fGE0}), we find (for real $p \ge 0$)
\begin{eqnarray}
&&\Re G_{E}^{0}(0,0;0,0)=\frac{m}{\hbar^{2}}\Biggl [
-\frac{1}{4\pi}\ln y_{c}
%\nonumber\\
+\frac{\kappa^{2}}{\kappa^{2}+p^{2}}
\Bigl (1-\frac{1}{\pi}{\rm arctan}\sqrt{\frac{W}{\epsilon_{b}}}\Bigr 
)\nonumber\\
&&~~~+\frac{\kappa p}{\kappa^{2}+p^{2}}\frac{1}{2\pi}\Bigl (
\ln\frac{\sqrt{W}+\sqrt{E+\epsilon_{b}}}{\sqrt{W}-\sqrt{E+\epsilon_{b}}}
%\nonumber\\
-\ln \frac{|y_{c}-y_{1}|}{|1-y_{c}y_{1}|}
\Bigr )\Biggr ],
\label{ReG}
\end{eqnarray}
with
%\begin{eqnarray}
$y_{c}=\Bigl (2\sqrt{W(W-E)}+2W-E\Bigr )^{2}/E^{2}$,
%, \nonumber\\
$y_{1}=(p+\kappa)^{2}/(p-\kappa)^{2}$,
%\end{eqnarray}
and
\begin{equation}
\Im G_{E}^{0}(0,0;0,0)=-\frac{2m}{\hbar^{2}}\Bigl
[\frac{p\kappa}{p^{2}+\kappa^{2}}
%\nonumber\\
+\Theta (E)\Bigl
(\frac{p-\kappa}{2}\Bigr )^{2}\frac{1}{p^{2}+\kappa^{2}}\Bigr ],
\label{ImG}
\end{equation}
%Here, we have used $E=E(p,b)=\hbar^{2}(p^{2}-\kappa^{2})/2m$, and
where $\Theta(E)$ is the Heavyside function.

When $p=0$, then $G_E^{0}(0,0;0,0)$ is real and the
equation $\Re G_{E}^{0}(0,0;0,0)=1/U$ (equivalent to    
the resonance $U_{\rm eff}(E)^{-1}=0$)
is identical to the equation for the transition from a doubly bound
ground state to a ground state with one `free' electron.
%eigenvalues of the interacting case
\cite{aharony}
In our special case, $\Re G_{E}^{0}(0,0;0,0)$ starts at a negative value
at $\kappa=0$, changes to positive values around $\kappa \approx 0.3$,
and approaches the asymptotic value $1/2$ at large $\kappa$.
Thus, the above equation has a solution only for $U>2$, when such a 
transition
occurs.
For small $\kappa$, $\Re G_{E}^{0}(0,0;0,0)$,
remains negative,
$U_{\rm eff}(E)$ remains repulsive and there
is no ``resonance"; the transmission increases monotonically with $U$,
as shown in Fig. \ref{fig22}(a).
For large $\kappa$ (and
$p>0$), 
$U_{\rm eff}(E)$ becomes {\it negative}, reflecting an {\it attractive}
effective interaction!
An explicit calculation now yields a peak in $T(U,p)$ just before
$U_{\rm eff}(E)$
changes sign, see Fig. \ref{fig22}(b).
These results are understandable qualitatively: at small $U$, the presence
of the ``bound" electron on the ``dot"  weakens
the attractive potential $V$, and thus causes an increase in the 
transmission. 
Indeed, a Hartree--Fock--like
approximation, in which we calculate the average of the interaction term
with the symmetrized wave function $\Psi^{pb}_0$, yields
$V \rightarrow V-2U\kappa$,
and thus
$\kappa \rightarrow \kappa
(1-2mU/\hbar^{2})$.
Using this renormalized value of $\kappa$ in the bare
transmission $|t_{p}|^{2}=p^{2}/(p^{2}+\kappa^{2})$, reproduces the result
(\ref{T}) in the small $U$ limit. Thus, 
the interaction $U$ compensates the  (single--particle)
attraction, and causes an increase in $T(U,p)$. \cite{pichard}
When $U_{\rm eff}(E)$
%the effective interaction 
changes sign and becomes 
negative,
the transmission has a pronounced peak.
A similar dramatic behavior of $T(U,p)$ is manifested by its
dependence on the depth of the single--particle bound state, $V$, or --
equivalently -- on $\kappa$, as shown in Fig. \ref{fig11}.
%In the absence of the
%interaction, the transmission decays as $\kappa $ increases, 
%$T(p)=|t_p|^2=p^2/(p^2+\kappa^2)$. 

%For a sufficiently localized electron, a large $U$ may turn the effective
%potential well into a potential {\it barrier}, and the transmission 
%decreases.

The above results indicate that the more interesting behavior of the 
transmission
occurs when the single--particle attraction is high enough, 
$\kappa > \omega$ (that is,
when the bound state energy $|\epsilon_{b} |$ is larger than the band 
width,
$W$).
In this situation, the total energy of the two electrons is negative 
[$p^2 < \omega^2 < \kappa^2$], and $\Im G_E^0$ is given only by the first 
term
in Eq. (\ref{ImG}).
Then, the transmission takes an especially simple form
\begin{equation}
T(U,p)=|\tilde{t}_{p}|^{2},~~~ \tilde{t}_{p}=t_{p}\Bigl (1-F_{E}r_{p}\Bigr 
)
=t_{p}\Bigl (\frac{1}{U_{\rm eff}}+|r_{p}|^{2}\Bigr )\Big /
\Bigl (\frac{1}{U_{\rm eff}}
+i\frac{p\kappa}{p^{2}+\kappa^{2}}\Bigr ).
\label{ttil}
\end{equation}
Hence, when $1/U_{\rm eff}$ approaches $0$, $\tilde{t}_{p}\rightarrow 
-r_{p}$.
At large values of $\kappa $ the reflection is close to unity. This 
explains the
heights of the peaks in Figs. \ref{fig22}(b) and \ref{fig11}.
%Indeed, using Eq. (\ref{ttil}), and the analogous result for the `dressed'
%reflection amplitude, $\tilde{r}_{p}=r_{p}(1-F_{E}t_{p})$, one obtains an
%expression for the effective inverse localization length of the bound 
%state, 
%\begin{eqnarray}
%\kappa_{\rm eff}
%=\kappa\Bigl (1-\frac{U_{\rm eff}}{1+|r_{p}|^{2}U_{\rm eff}}\Bigr ).
%\end{eqnarray}

Returning to Eq. (\ref{ReG}), is is interesting to note that the sign
change in $\Re G_E^0$ originates from the last two terms, which vanish
when $\kappa=0$. In the absence of the bound state, $U_{\rm eff}$
always remains negative. In the usual theory of superconductivity,
the remaining first term in Eq. (\ref{ReG}) renormalizes $U$ into
a weaker repulsion, crucial for the superconducting state. All the
interesting phenomena found here result from the {\it additional}
effects of the bound state.

Finally, a comment about spins: starting with two general spins
$\sigma_1$ and $\sigma_2$, one can always split the wave function into
a combination of a singlet and a triplet. Without the interaction,
the scattered electrons will have the same spins. Given the above results,
the interaction will change the relative weights of the singlet and the
triplet, and thus may cause a spin flip.
%, for the case when initially one electron was bound.

%\stars

This project is supported by the Israel Science Foundation
and by a joint grant from the Israeli Ministry of Science
and the French Ministry of Research and Technology.
O. E-W also thanks the Albert Einstein Minerva Center for Theoretical 
Physics
for partial support.

\vspace*{-4mm}
\newcommand{\noopsort}[1]{} \newcommand{\printfirst}[2]{#1}
  \newcommand{\singleletter}[1]{#1} \newcommand{\switchargs}[2]{#2#1}

%\end{multicols}{2}

%\end{document}

%\end{multicols}{2}
%\vspace{-0.5cm}
\begin{figure}
%\narrowtext
%\vspace{1cm}
\centerline{\psfig{figure=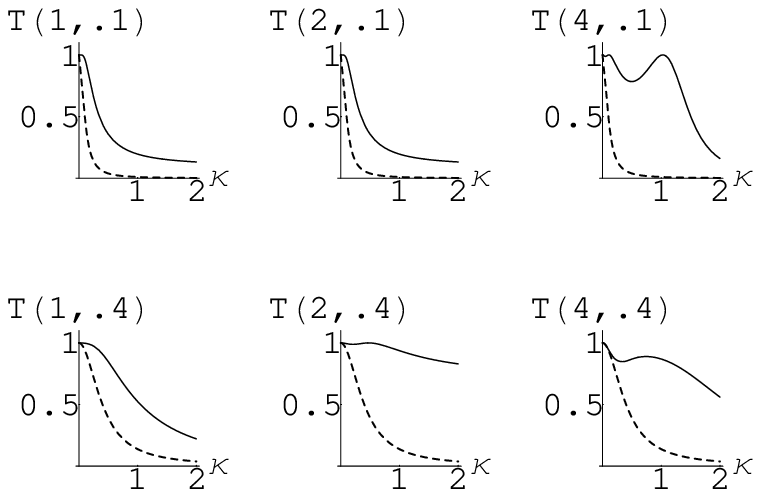,width=14cm}}
\vspace{-14cm}
\caption{The
transmission $T(U,p)$
as function of $\kappa $ (in units of the cutoff $\omega$),
for $U=1,~2,~4$ (in units of $\hbar^2/(2m)$), solid line,
and
$T(0,p)=|t_{p}|^{2}$, dashed line.
$p$ is in units of $\omega$.}
\label{fig11}
\end{figure}
%\vspace{-0.5cm}

%\vspace{-8cm}
\begin{figure}
%\narrowtext
%\vspace{1cm}
\centerline{\psfig{figure=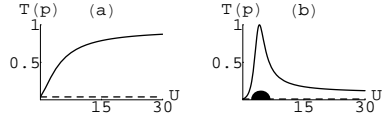,width=14cm}}
\vspace{-15cm}
\caption{The
transmission $T(U,p)$ as function of the `bare' interaction $U$ (in units
of $\hbar^2/(2m)$).
The solid line is
$T(U,p)$, the dashed line shows $|t_{p}|^{2}$. (a) $\Re G_{E}^{0}<0$
($\kappa =0.2\omega$, $p=0.04\omega$). (b) $\Re G_{E}^{0}>0$ ($\kappa
=\omega$,
$p=0.1\omega$). The thick point signifies the change of sign of $U_{\rm
eff}$.}
\label{fig22}
\end{figure}

\end{document}